\documentclass{article}       % onecolumn (second format)

\usepackage{graphicx}
\usepackage{array}
\usepackage{hyperref} % links

\usepackage{tabularx} % even nicer tables
\usepackage{booktabs}
\usepackage{amsmath}
\usepackage{xcolor}

\newcolumntype{Y}{>{\centering\arraybackslash}X}
\newcolumntype{R}{>{\raggedleft\arraybackslash}X}%
\usepackage{listings}

\begin{document}

\title{Accelerated Computing in Magnetic Resonance Imaging -- Real-Time Imaging Using Non-Linear Inverse Reconstruction}
\author{Sebastian Schaetz,$^1$  Dirk Voit,$^1$ Jens Frahm,$^{1,3}$ and Martin Uecker$^{2,3}$}

\maketitle

\begin{abstract}
\textbf{Purpose:}
To develop generic optimization strategies for image
reconstruction using graphical processing units (GPUs)
in magnetic resonance imaging (MRI) and to exemplarily
report about our experience with a highly accelerated implementation
of the non-linear inversion algorithm (NLINV) for
dynamic MRI with high frame rates.
~\\
\textbf{Methods:}
The NLINV algorithm is optimized and ported to run on an a multi-GPU single-node server.
The algorithm is mapped to multiple GPUs by decomposing the data domain along the channel dimension. Furthermore, the algorithm is decomposed along the temporal domain by relaxing a temporal regularization constraint, allowing the algorithm to work on multiple frames in parallel. Finally, an autotuning method is presented that is capable of combining different decomposition variants to achieve optimal algorithm performance in different imaging scenarios.
~\\
\textbf{Results:}
The algorithm is successfully ported to a multi-GPU system and allows online image reconstruction with high frame rates.
Real-time reconstruction with low latency and frame rates up to 30 frames per second is demonstrated.
~\\
\textbf{Conclusion:}
Novel parallel decomposition methods are presented which are applicable to many
iterative algorithms for dynamic MRI. Using these methods to parallelize the NLINV
algorithm on multiple GPUs it is possible
to achieve online image reconstruction with high frame rates.

~

\end{abstract}

\footnotetext[1]{Biomedizinische NMR Forschungs GmbH at the Max Planck Institute for Biophysical Chemistry, G\"ottingen, Germany}
\footnotetext[2]{Department of Diagnostic and Interventional Radiology, University Medical Center G\"ottingen, G\"ottingen, Germany}
\footnotetext[3]{DZHK (German Centre for Cardiovascular Research), Partner Site G\"ottingen, Germany}

\newpage

\section{Introduction}

Accelerators such as graphics processing units (GPUs) or other multi-core vector co-processors are well-suited for achieving fast algorithm run times when applied to reconstruction problems in medical imaging \cite{eklund2013medical} including computed tomography \cite{scherl2007fast}, positron emission tomography \cite{schaetz2010integration} and ultrasound \cite{dai2010real}. This is because respective algorithms usually apply a massive number of the same independent operations on pixels, voxels, bins or sampling points. The situation maps well to many-core vector co-processors and their large number of wide floating-point units that are capable of executing operations on wide vectors in parallel. The memory bandwidth of accelerators is roughly one order of magnitude greater than that of central processing units - they are optimized for parallel throughput instead of latency of a single instruction {stream.}

{Recent} advances in Magnetic Resonance Imaging (MRI) pose new challenges to the implementations of associated 
algorithms in order to keep data acquisition and reconstruction times on par. While acquisition times decrease 
dramatically as in real-time MRI, at the same time the data grows in size when using up to 64 or 
even 128 independent receiver channels on modern MRI systems. In addition, new modalities such as model-based reconstructions for (dynamic) parametric mapping increase the computational complexity of reconstruction algorithms because they in general involve iterative solutions to non-linear inverse problems. As a consequence, accelerators are increasingly used to overcome these challenges \cite{stone2008accelerating,hansen2008cartesian,gai2013more,sorensen2009real,uecker2010nonlinear,schaetz2012,murphy2012fast,smith2012cs,uecker2015bart}.

\begin{figure}
\centering
	\includegraphics[width=80mm]{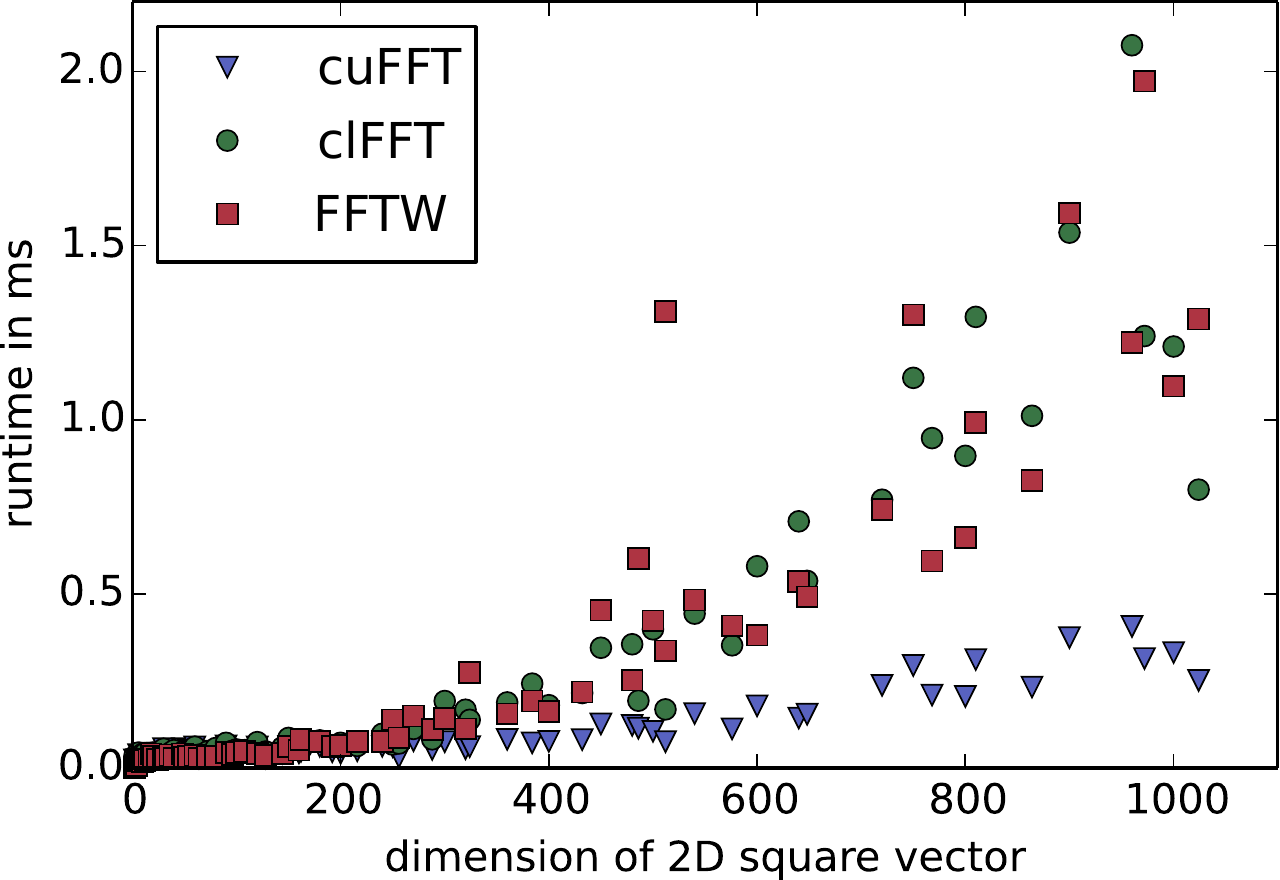} 
	\caption{Run time (i.e., minimal wall-clock time in ms) measurement of a single-precision 2D squared complex-to-complex out-of-place Fourier transform of FFTW, clFFT and cuFFT libraries. clFFT and cuFFT benchmarks were obtained for NVIDIA GeForce Titan Black, the FFTW benchmark was obtained for Intel Xeon E5-2650 utilizing 8 threads. FFTW is initialized with \lstinline|FFTW_MEASURE|. The clFFT library has only limited support for mixed radix FFTs and achieves good performance for power of 2 FFTs only.}
	\label{fig:fft-case-plot}
\end{figure}

\begin{figure}
\centering
	\includegraphics[width=80mm]{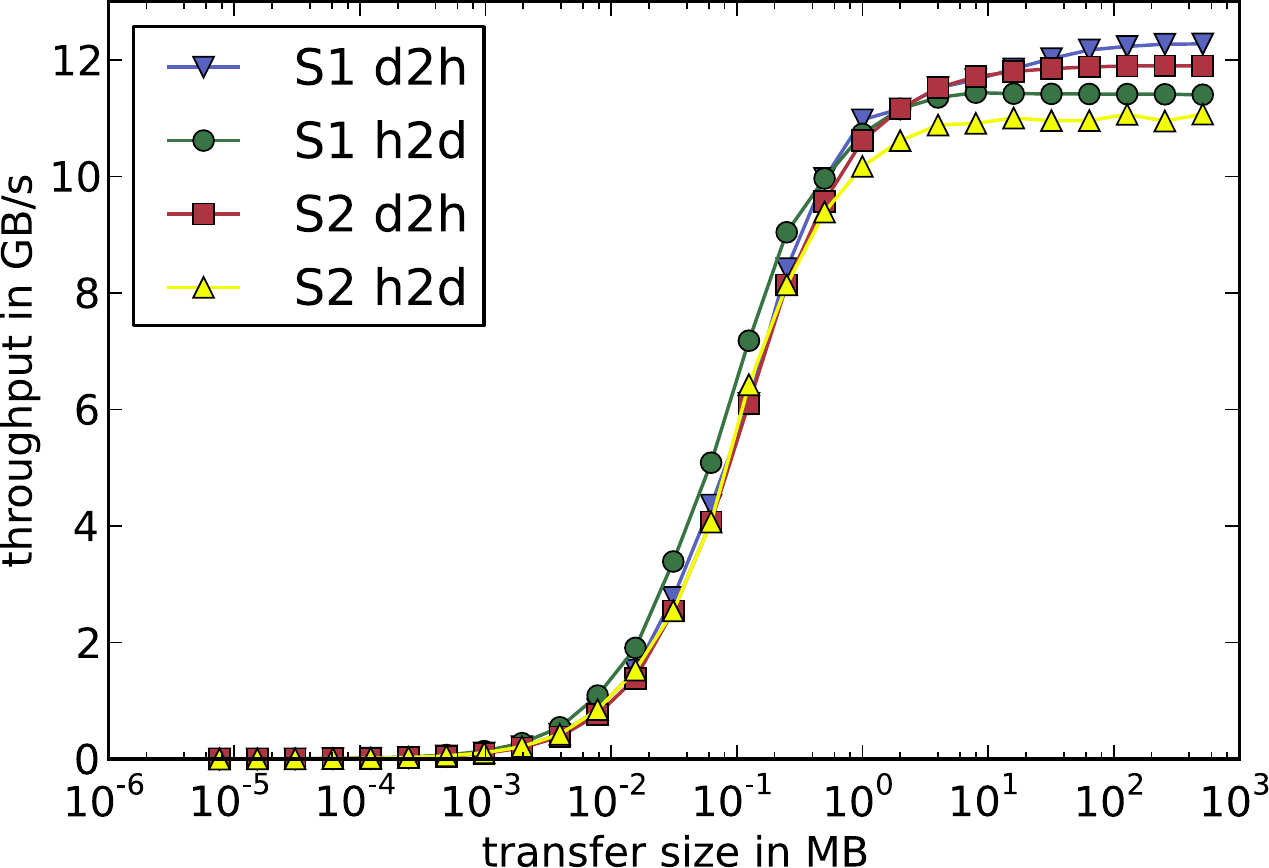} 
	\caption{Throughput of Supermicro (S1) and Tyan (S2) 8x PCIe 3.0 systems: Device-to-host and
	host-to-device transfer of non-pageable (pinned) memory with junk sizes between 8 bytes and 1 GB}
	\label{fig:throughput}
\end{figure}

Accelerators outperform main processors (CPUs) in all important operations during MRI reconstruction 
including interpolation \cite{gregerson2008implementing}, filtering \cite{huang2009non} and basic linear algebra \cite{volkov2008benchmarking}. The central operation of most MRI reconstruction algorithms is the Fourier 
transform. Figure \ref{fig:fft-case-plot} shows the performance of the three FFT libraries cuFFT, clFFT and 
FFTW \cite{frigo1998fftw}. The accelerator libraries cuFFT and clFFT outperform the CPU 
library FFTW. 
Consequently, accelerators outperform CPUs for MRI
reconstruction, if the algorithm is based on the Fourier transform and if the ratio of computation
to data transfer favours computation.
{
Accelerator and CPU typically form distributed memory systems. 
Figure \ref{fig:throughput} shows the memory transfer speed for two different PCIe 3.0 systems 
in dependence of the data transfer size. The amount of computation assigned to an accelerator and 
the transfer size have a huge impact on the overall performance of the algorithm, which
needs to be taken into account when implementing accelerated algorithms for image reconstruction.}
Another limiting factor for the use of accelerators in MRI is the restricted amount of memory available. Nevertheless, the on-board memory of accelerators increased in recent years and the newest generation of accelerators is equipped with acceptable amounts of memory (compare Table \ref{tbl:acc_memory_amount}).

\begin{table}
{
\caption{Memory, memory-bandwidth, and single-precision performance
of high performance computing (HPC) and consumer accelerators (source: Wikipedia)
%\url{https://en.wikipedia.org/wiki/List_of_Nvidia_graphics_processing_units}
%\url{https://en.wikipedia.org/wiki/AMD_Radeon_Rx_200_series}
%\url{https://en.wikipedia.org/wiki/Xeon_Phi}
}\label{tbl:acc_memory_amount}
\centering
\begin{tabularx}{\columnwidth}{l|X|c|r|r|r|r}
\toprule
Vendor  & Model          & HPC & Memory & Bandwidth & Performance\\
\midrule
NVIDIA & K40            & yes & 12 GB & 288 GB/s & 4.29 TFLOPS  \\
NVIDIA & Titan Black    & no  & 6 GB & 336 GB/s & 5.12 TFLOPS \\
AMD    & FirePro W9100  & yes & 16 GB & 320 GB/s & 5.24 TFLOPS \\
AMD    & Radeon R9 290X & no  & 6 GB & 320 GB/s & 5.63 TFLOPS \\
Intel  & Xeon Phi 7120  & yes & 16 GB & 352 GB/s & 2.40 TFLOPS \\
\bottomrule
\end{tabularx}
}
%NVIDIA & Titan Xp 	& no  & 12 GB & 547.7 & 3840 & (1582) \\
\end{table}

{In this paper, we summarize our experience in developing
a low-latency online reconstruction system for real-time MRI over
the last eight years. While some of the described techniques have already
been reported previously, this article for the first time explains all
technical aspects of the complete multi-GPU implementation of the advanced
iterative reconstruction algorithm, and adds further background information
and analysis. Although we focuss on the specific application of real-time
MRI, many of the techniques developed for this project can be applied to
similar tomographic reconstruction problems. In particular, we describe
strategies for optimal choice of the grid size for a convolution-based
non-uniform FFT, novel parallelization schemes using temporal and spatial
decomposition, and automatic tuning of parameters.}

\section{Theory}
\subsection{Real-Time MRI}

Diagnostic imaging in real time represents a most demanding acquisition and reconstruction problem 
for  MRI. In principle, the data acquisition refers to the recording of a large number of different radiofrequency signals in the time domain which are spatially encoded with the use of magnetic field gradients. The resulting dataset represents the k-space (or Fourier space) of the image.  MRI acquisition times are determined by the number of different encodings needed for high-quality image reconstruction multiplied by the time required for recording a single MRI signal, i.e. the so-called repetition time TR. While TR values could efficiently be reduced from seconds to milliseconds by the invention of low-flip angle gradient-echo MRI sequences, e.g. see \cite{frahm1986rapid} for an early dynamic application, a further speed-up by reducing the number of encodings was limited by the properties of the Fourier transform which for insufficient coverage of k-space causes image blurring and/or aliasing artifacts. Reliable and robust improvements in acquisition speed by typically a factor of two were first achieved when parallel MRI techniques \cite{sodickson1997simultaneous} \cite{pruessmann1999sense} where introduced. These methods compensate for the loss of spatial information due to data undersampling by simultaneously acquiring multiple datasets with different receiver coils. In fact, when such multi-coil arrangements are positioned around the desired field-of-view, e.g. a head or thorax, each coil provides a dataset with a unique spatial sensitivity profile and thus complementary information. This redundancy may be exploited to recover the image from moderately undersampled k-space data and thus accelerate the scan. 

Parallel MRI indeed was the first concept which changed the MRI reconstruction from a two-dimensional FFT to the solution of an inverse problem. To keep mathematics simple and computations fast, however, commercially available implementations turned the true nonlinear inverse problem, which emerges because the signal model contains the product of the desired complex image and all coil sensitivity profiles (i.e., complex images in themselves), into a linear inverse problem. This is accomplished by first determining the coil sensitivities with the use of a pre-scan or by performing a low-resolution Fourier transform reconstruction of acquisitions with full sampling in the center of k-space and undersampling only in outer regions.

\begin{figure}
\centering
	\includegraphics[width=80mm]{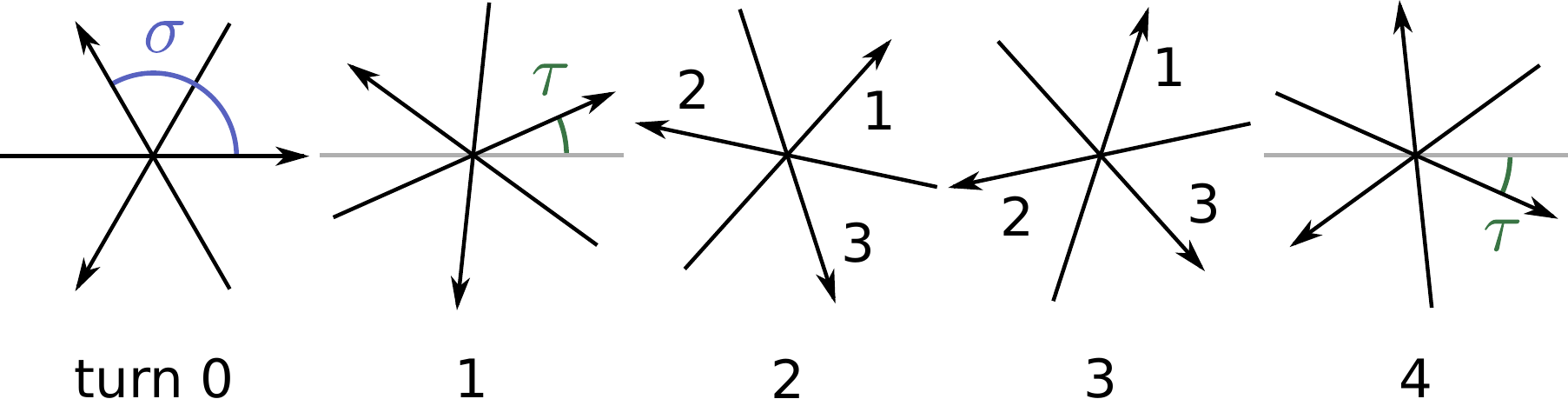} 
	\caption{Schematic acquisition scheme for real-time MRI with $U = 5$ different sets of spokes, $K = 3$, $\sigma = 2\pi/K$ and $\tau = 2\pi/(KU)$.}
	\label{fig:reorder}
\end{figure}

Recent advances towards real-time MRI with so far unsurpassed spatiotemporal resolution and image quality \cite{uecker2010real} were therefore only possible by combining suitable acquisition techniques with an adequate image reconstruction. Crucial elements include the use of (i) rapid low-flip angle gradient-echo sequences, (ii) spatial encodings with radial rather than Cartesian trajectories, (iii) different (i.e., complementary) sets of spokes in successive frames of a dynamic acquisition (see Figure \ref{fig:reorder}) \cite{zhang2010magnetic}, (iv) extreme data undersampling for each frame, (v) an image reconstruction algorithm (NLINV) that solves the non-linear inverse problem (see below), and (vi) temporal regularization to the preceding frame which constrains the ill-conditioned numerical problem with physically plausible a priori knowledge. Real-time MRI with NLINV reconstruction achieves a temporal resolution of 10 to 40 ms per frame (i.e., 25 to 100 frames per second) depending on the actual application (e.g., anatomic imaging at different spatial resolutions or quantitative blood flow studies), for a recent review of cardiovascular applications see \cite{zhang2014real}.

The spatial encoding scheme is comprised of $U$ different sets (turns) of $K$ spokes. All $U$ sets of spokes taken together 
cover the k-space uniformly. Figure \ref{fig:reorder} illustrates the real-time MRI encoding scheme.

\subsection{Image Reconstruction with NLINV}

If the receive coil sensitivities in parallel MRI are known, the image recovery emerges as a linear inverse problem which can efficiently be solved using iterative methods. In practice, however, static sensitivities are obtained through extrapolation and, even more importantly, in a dynamic in vivo setting they change due to coupling with the conductive tissue: this situation applies to a human subject during any type of movement (e.g., breathing or cardiac-related processes) or when dynamically scanning different planes and orientations (e.g., during real-time monitoring of minimally invasive procedures). 

In such situations, extrapolating static sensitivities is not sufficient. The sensitivities have to be jointly estimated together with the image, yielding an inverse reconstruction problem that is non-linear and ill-posed. For real-time MRI, both the dynamic changes during a physiologic process and the need for extremely undersampled data sets (e.g., by a factor of 20) inevitably lead to a nonlinear inverse problem.

%\subsubsection{NLINV Algorithm}

A powerful solution to this problem is the regularized nonlinear inverse reconstruction 
algorithm~\cite{uecker2008image}. NLINV formulates the image reconstruction as a nonlinear least-squares problem
\begin{align}
	\operatorname{argmin}_x \underbrace{\| Fx - y \|_2^2}_{\textrm{data fidelity}} 
		+ \underbrace{\alpha \|W( x - x_{prev} ) \|_2^2}_{\textrm{regularization}}~.
\end{align}
The nonlinear forward operator $F = \mathcal{F} \circ C$ maps the combined vector $x$ of the unknown complex-valued image $\rho$ and all coil sensitivities $c_j$ (i.e., complex-valued images in themselves) to the data. $C$ multiplies the image with the sensitivities to obtain individual coil images $c_j \rho$ and $\mathcal{F}$ then predicts the k-space data using a non-uniform Fourier transform for all coil elements $j = 1, \dots, J$ \footnote{This non-uniform FFT is implemented here with a uniform FFT and convolution with the {PSF\cite{wajer2001major}, which is especially advantageous for implementation on a GPU~\cite{uecker2010nonlinear}.}}. The data fidelity term quantifies the difference between this predicted data and the measured data in the least-squares sense, while the additional regularization term addresses the ill-posedness of the reconstruction problem. Here, $W$ is a weighting matrix in the Fourier domain which characterizes the spatial smoothness of the coil sensitivities and does not change the image component of $x$. For dynamic imaging, temporal regularization to the immediately preceding frame $x_{prev}$, which exploits the temporal continuity of a human movement or physiologic process, allows for a remarkably high degree of data undersampling or image acceleration \cite{uecker2010real}.

\begin{figure}
	\includegraphics[width=\columnwidth]{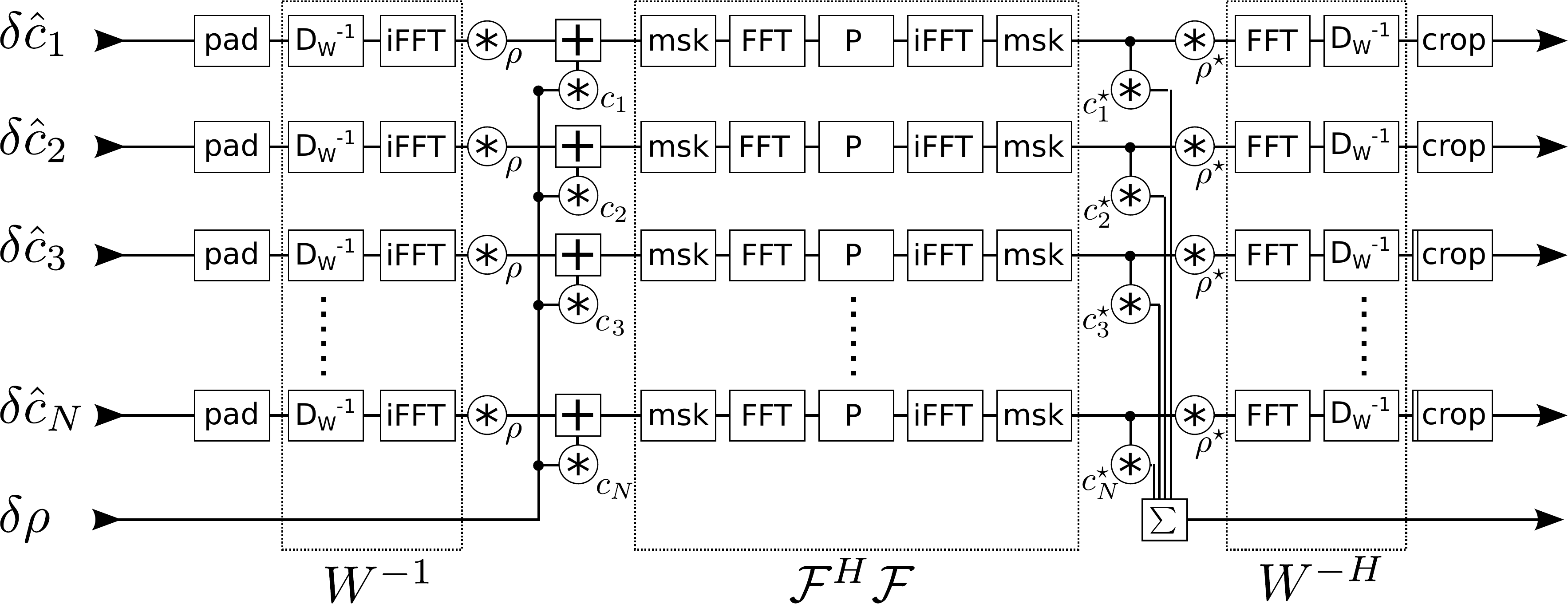}
	\caption{Flowchart of the operator $D{\hat F}^H_{\hat x_n} D{\hat F}_{\hat x_n}$. The
	implementation of the weighting matrix $W^{-1}$ and its adjoint $W^{-H}$ apply a diagonal weighting
	matrix $D_W^{-1}$ and a forward or inverse fast Fourier transform (FFT, iFFT) for each channel.
	The derivative of the multiplication of the image $\rho$ and the coil sensitivities $c_j$
	and its adjoint are composed of point-wise multiplication ($\star$), addition ($+$), and summation ($\sum$)
	operations. The combination of the non-uniform Fourier transform and its adjoint
	$\mathcal{F^H} \mathcal{F}$ is implemented as a convolution with the point-spread function,
	consisting of a mask (msk) restricting the oversampled grid to the field-of-view,
	application of fast Fourier transform and its inverse,
	and a point-wise multiplication with the convolution kernel (P). Padding (pad) and
	cropping (crop) operations are used to reduce computational cost in the overall
	iterative algorithm.}
	\label{fig:flow}
\end{figure}

In the following, the main steps of an efficient numerical implementation are described, for further details see \cite{uecker2010real} \cite{uecker2010nonlinear}. First, a change of variables
$\hat x = W x$ and $\hat F = F \circ W^{-1}$
is done to improve the conditioning of the reconstruction problem. Starting from an initial guess $x_0$, the numerical problem is then solved using the iteratively regularized Gauss-Newton method (IRGNM), which uses the following linear update rule
\begin{align}
	&\left( D{\hat F}^H_{\hat x_n} D{\hat F}_{\hat x_n} + \alpha_n I \right) (\hat x_{n+1} - \hat x_n) \\
	&\quad = D{\hat F}^H_{\hat x_n} (y - {\hat F} \hat x_n) - \alpha_n (\hat x_n - \hat x_{prev})~.
\end{align}
Here, $D{\hat F}_{\hat x_n}$ denotes the derivative of $\hat F$ at $\hat x_n$ and $D{\hat F}^H$ its adjoint.
In each of the $M$ Newton steps, this linear system of equations is solved using the method of conjugate gradients (CG). The main operation in a matrix-free implementation is the repeated application of the operator
\begin{align}
D{\hat F}^H_{\hat x_n} D{\hat F}_{\hat x_n}: \delta x \mapsto \left( \begin{array}{c} \sum_{j=1}^J c_j^{\star} t_j \\
					W_b^{-H} \rho^{\star} t_1  \\
					\vdots \\
					W_b^{-H} \rho^{\star} t_J \end{array}\right) \quad
\end{align}
with 
\begin{align}
t_j := \mathcal{F}^H_b \mathcal{F}_b \left\{ c_j \delta \rho + \rho W_b^{-1} \delta \hat c_j \right\} .
\end{align}
The star $\star$ means complex conjugation. $\mathcal{F}_b$ and $W_b$ are the blocks of the block-diagonal operations  $\mathcal{F}$ and $W$ which operate on a single channel. Because the non-uniform Fourier transform is paired with its adjoint, it can be implemented efficiently as a truncated convolution with a point-spread function using two applications of an FFT algorithm on a twofold oversampled grid \cite{wajer2001major}. $W$ is implemented using one FFT followed by the application of a diagonal weighting matrix. Thus, a single iteration requires 4 applications of an FFT algorithm per channel, several pixel-wise complex multiplications and additions, and one pixel-by-pixel reduction across all channels ($\sum_{j=1}^J$). A flowchart can be found in Figure~\ref{fig:flow}.

The NLINV algorithm is an iterative algorithm consisting of operators $F$, $D{\hat F}$ and $D{\hat F}^H$ as well as the conjugate gradient method. Depending on the imaging scenario, the number of Newton-steps are fixed to $6$ to $10$. $F$ is only computed once per Newton-step. $D{\hat F}$ and $D{\hat F}^H$ are applied in each conjugate-gradient iteration. Each operator applies a 2D FFT twice on per-channel data. $F$ computes a scalar product once, the conjugate gradient method contains two scalar products over all channel data. Furthermore each operator applies between $4$ to $6$ element-wise operations and $D{\hat F}^H$ contains a summation over channel data. The performance of the Fourier transform dominates the run-time of a single inner-loop iteration of the algorithm and it has to be applied $4$ times in each inner-loop. 

If $10$ channels are assumed, $6$ Newton steps are computed (yielding roughly 50 conjugate gradient iterations) $10*4*50 = 2000$ 2D Fourier transformations must be applied to compute a single image. If data is acquired at a frame-rate of $30$ fps, the reconstruction system must be able to apply $60000$ 2D Fourier transforms per second.

\section{Optimization Methods and Results}

The following section outlines several steps undertaken to move a prototype implementation of the NLINV algorithm to a highly accelerated implementation for use in an online reconstruction pipeline. The original implementation in C utilizes the CUDA framework and is capable of reconstructing at speeds of 1 to 5 fps on a single GPU. A typical clinical scenario in real-time MRI is cardiac imaging where a rate of 30 per second is necessary. To incorporate real-time heart examination in a clinical work-flow, the image reconstruction algorithm must perform accordingly.

The optimization techniques are classified into two categories. The first category encompasses platform independent optimization procedures that reduce the overall computational cost of the algorithm by reducing vector sizes while maintaining image quality. In the second category, the algorithm is modified to take advantage of the multi-GPU computer platform by channel decomposition, temporal decomposition and tuning of the two techniques to yield best performance for a given imaging scenario. 

All benchmark results show the minimum wall-clock time of a number of runs. For micro-benchmarks hundreds of runs were performed. For full reconstruction benchmarks, tens of runs were performed. The speed-up $S$ is defined as the quotient of the old wall-clock time $t_{old}$ over the new time $t_{new}$ 
\begin{equation}
S = \frac{t_{old}}{t_{new}}
\end{equation}
and parallel efficiency $E$ is defined as
\begin{equation}
E = \frac{S_p}{p}
\end{equation}
with $S_p$ representing the speed-up achieved when using $p$ accelerators over $1$. Perfect efficiency is achieved when $S_p = p$ or $E = 1$, sublinear speed-up is measured if $S_p < p$ and $E<1$.

If not stated otherwise, all benchmarks were obtained on a Supermicro SuperServer 4027GR-TR system with PCIe 3.0, 2x Intel Xeon Ivy Bridge-EP E5-2650 main processors, 8x NVIDIA Titan Black (Kepler GK110) accelerators with 6 GB graphics memory each, and 128 GB main system memory. Custom GPU power cables were fabricated for the Supermicro system to support consumer-grade accelerators. The CUDA run time environment version was 6.5 with GPU driver version 346.46. The operating system used was Ubuntu 14.04. 
{The combined single-precision performance of the eight GPUs in this system approaches 41 TFLOPS and the combined memory bandwidth is 2688 Gb/s.
For a comparison of the performance of some accelerators available at time of writing of the article, see Table \ref{tbl:acc_memory_amount}.}

\subsection{Real-time Pipeline}

The  real-time NLINV reconstruction algorithm is part of a larger signal processing pipeline that can be decomposed into several stages:
\begin{itemize}
	\item datasource: reading of input data into memory
	\item preprocessing: interpolating data acquired with a non-Cartesian trajectory onto a rectangular grid, correcting for gradient delays as well as performing the channel compression; this stage contains a calibration phase that calculates the channel compression transform matrix and estimates the gradient delay
	\item reconstruction: the NLINV algorithm, reconstructing each image
	\item postprocessing: cropping of the image to the measured field-of-view, calculation of phase-difference image (in case of phase-contrast flow MRI), applying temporal and spatial filters
	\item datasink: writing of resulting image to output
\end{itemize}

Real-time MRI datasets with high temporal resolution typically contain several hundreds of frames even when covering only a few seconds of imaging. For this reason it is beneficial to parallelize the pipeline at the level of these functional stages: each stage can work on a different frame, or in other words, different frames are processed by different stages of the pipeline in parallel. The implementation of such a pipeline follows the actor model \cite{greif1975semantics} where each pipeline stage is one (or multiple, see \ref{subsec:temp_decomp} Temporal decomposition) actor, receiving input data from the previous stage as message and passing results to the following stage as a different message. The first actor and the last actor only produce (datasource) and receive (datasink) messages, respectively.

\begin{figure}
\centering
	\includegraphics[width=80mm]{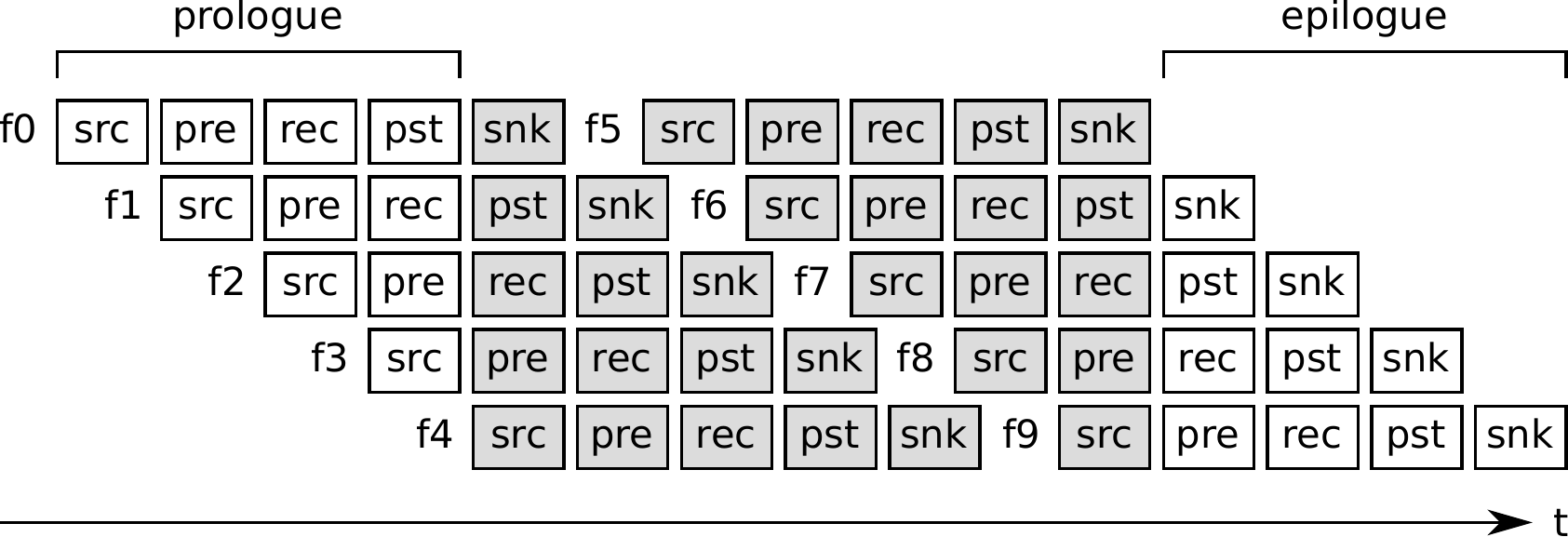} 
	\caption{The NLINV reconstruction pipeline consists of 5 stages: datasource (src), preprocessing (pre), reconstruction (rec), postprocessing (pst) and datasink (snk). New frames (f0 through f9) enter the pipeline through the datasource and leave the pipeline through the datasink. A filled pipeline processes 5 frames at the same time by using 5 threads, one for each pipeline stage.}
	\label{fig:pipeline}
\end{figure}

Figure \ref{fig:pipeline} shows the NLINV pipeline for reconstructing 10 frames. The pipeline has a prologue and an epilogue of 4 frames (number of pipeline stages minus one) where full parallel reconstruction is not possible because the pipeline is filling up or emptying.

\subsection{Computational Cost Reduction}

\begin{figure}
\centering
	\includegraphics[width=80mm]{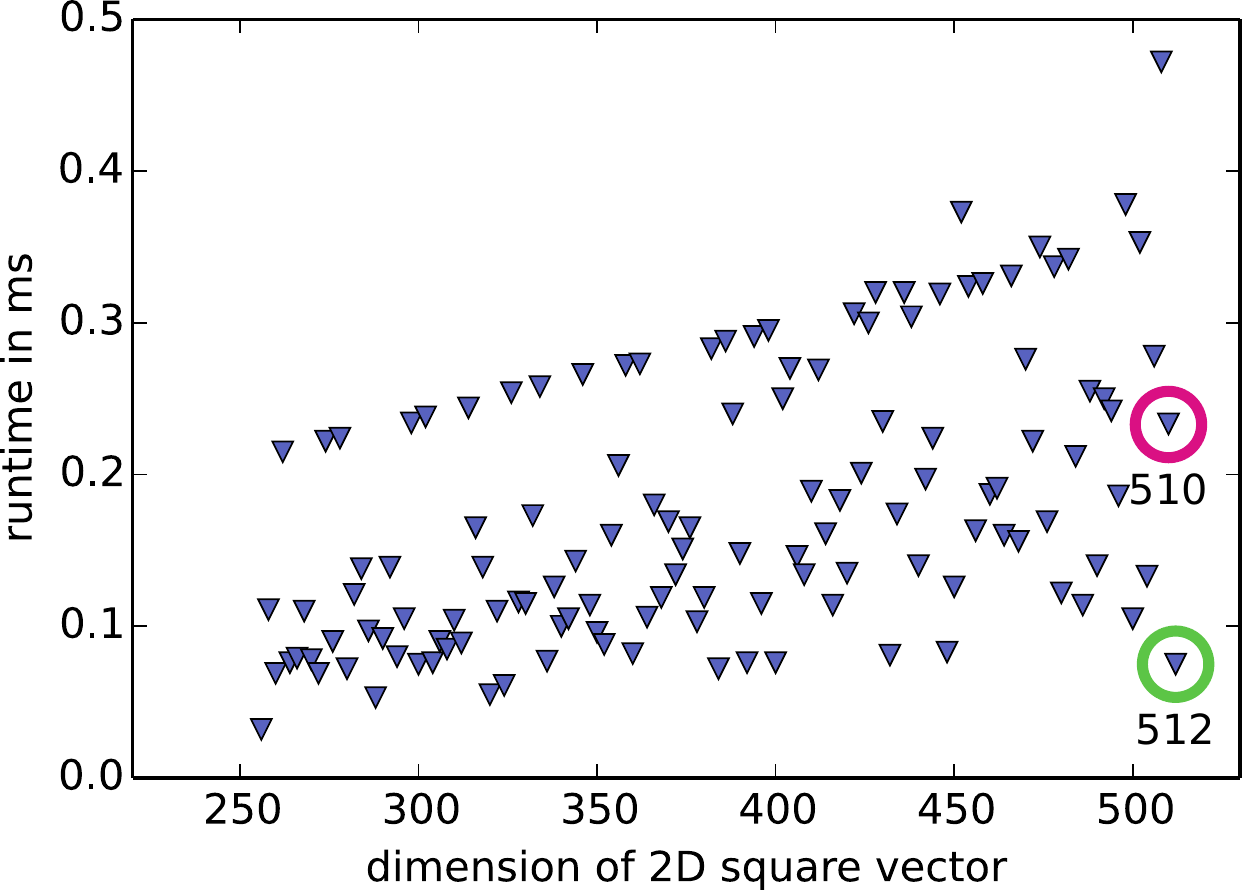} 
	\caption{Run time in ms of a single-precision 2D complex-to-complex out-of-place squared Fourier transform calculated using cuFFT. The plot shows run times as a function of vector size and highlights the run time difference between a $510^2$ vector (red) and $512^2$ vector (green). The plotted data ranges from $256^2$ to $512^2$.}
	\label{fig:tiny-plot}
\end{figure}

The  most compute-intensive part of the NLINV algorithm is the application of the Fourier transform. NLINV applies the Fourier transform multiple times in each iteration. Fast execution of the Fourier transform has thus a major impact on the overall run time of the algorithm. The FFT library cuFFT \cite{nvidia2010cufft} of the hardware vendor is used. Plotting the performance of the FFT libraries against the input vector size does not yield a linear function, but shows significant fluctuation (compare Figure \ref{fig:tiny-plot}). The algorithm performs better for vector sizes which are factorizable by small prime numbers, with the smallest prime numbers yielding the best results.

The grid oversampling ratio defines the ratio of grid sample locations $G$ over the side length of the output image $N$. The ratio between $N$ and $G$ includes an inherent factor of $2$ that exists as the convolution with the point-spread function requires twofold oversampling. An additional factor $\gamma$ can be set to values between $1$ and $2$ to prevent aliasing artifacts for situations where the operator choose a fields-of-view smaller than the object size. The relationship of $N$ to $G$ is thus defind as
\begin{equation}
G = 2 \gamma N.
\end{equation}
In the following, only $\gamma$ is specified. It is found that considering the constraints of the FFT algorithm when choosing $\gamma$ can yield calculation speed-ups. A lookup table is generated that maps grid size to FFT performance by benchmarking the Fourier transform algorithm for all relevant input sizes. This lookup table is generated for different accelerator generations as well as for new versions of the FFT library. The grid oversampling ratio is adjusted according to the highest measured FFT performance with a minimum oversampling ratio 
of $\gamma\geq1.4$. Table \ref{tbl:oversampling_ratio} shows the speed-up when comparing to a fixed oversampling ratio of $1.5$. Figure \ref{fig:tiny-plot} is a graphical representation of the lookup table that is used to select optimal grid sizes. It highlights the run time difference between two very similar vector sizes of $510^2$ ($2*3*5*17$) and $512^2$ ($2^9$). 

\begin{table}
\caption{Reconstruction times in fps for various image sizes with side length $N$ and fixed vs variable oversampling ratio $\gamma$. For some input sizes, $1.5$ is the optimal value ($N=128$, $N=144$), for other sizes, the number of grid sample locations must be increased to yield a speed-up ($N=160$, $N=170$). For a third group of image sizes ($N=256$), the optimal FFT size increases the size of the data to a degree that the speed-up is reduced or even nullified by additional overhead. For each test case 200 frames were reconstructed using 1 accelerator.}
\begin{tabularx}{\columnwidth}{Y|cr|lcr|Y}
\toprule
     & \multicolumn{2}{c|}{\textbf{fixed $\gamma=1.5$}} & \multicolumn{3}{c|}{\textbf{optimal $\gamma\geq1.4$}} &         \\
\midrule
$N$ & $G$        & fps         & $\gamma$             & $G$      & fps       & $S$ \\
\midrule
128  & 384         & 8.1        & 1.5           & 384       & 8.3      & 1.02    \\
144  & 432         & 2.5        & 1.5           & 432       & 2.5      & 1.00    \\
160  & 480         & 4.4        & 1.51875       & 486       & 5.0      & 1.25    \\
170  & 510         & 2.5        & 1.50588       & 512       & 5.8      & 2.32    \\
256  & 768         & 2.4         & 1.53125       & 784       & 2.4       & 1.00   \\
\bottomrule
\end{tabularx}
\label{tbl:oversampling_ratio}
\end{table}

\label{subsec:crop_pad}

\begin{figure}
\centering
	\includegraphics[width=80mm]{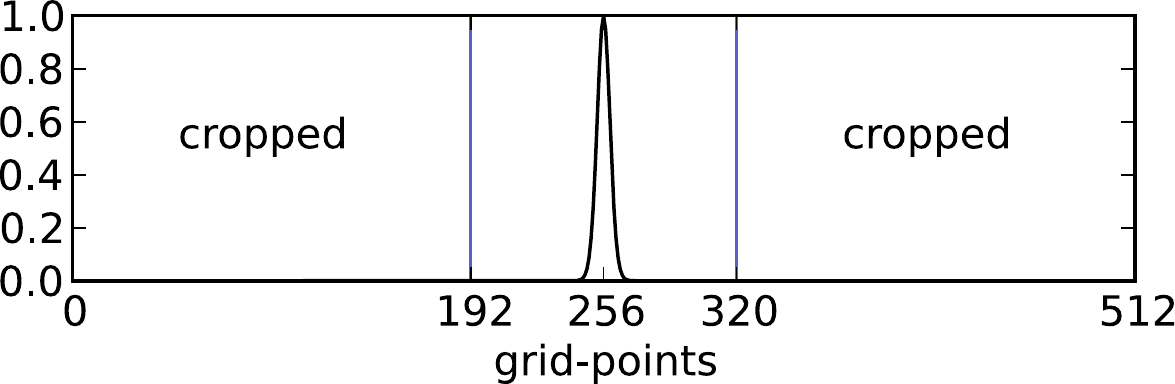} 
	\caption{The k-space weight function $(1+880|k|^2)^{16}$ with $-0.5<k_x,k_y<0.5$ is plotted in 1D for a grid-size of $512^2$ with cropping to 25\% of the original. The lowest frequency is at the center.}
	\label{fig:weight_func}
\end{figure}

The NLINV algorithm not only estimates the spin-density map $\rho$, but also the coil sensitivity profiles $c_j$. The regularization term added to the Gauss-Newton solver constrains the $c_j$ coil sensitivity profiles to a few low-frequency components. It is thus possible to not store the entire vector $c_j$, but to reduce its size to $\frac{1}{4}^2$ of the original. The number of grid sample locations $G$ is thus reduced to $G_c = \lfloor\frac{1}{4} G\rfloor$ for the coil sensitivity profiles $c_j$. Whenever the original size is required, the vector is padded with zeroes. Figure \ref{fig:weight_func} shows a typical weighting function in 1D as applied to $c_j$ and the cut-off that is not stored after this optimization. This saves computing time as functions applied on $c_j$ have to compute less. For this technique to yield a speed-up, the computation time saved must exceed the time required to crop and pad the vectors. Table \ref{tbl:tiny} shows the speed-up achieved by this optimization for 
various data sizes.

\begin{table}
\caption{Speed-up $S$ by cropping the 2D vector of size length $N$ of coil sensitivity profiles by a factor of $\frac{1}{4}^2$.}
\begin{tabularx}{\columnwidth}{Y|cr|cr|Y}
\toprule
     & \multicolumn{2}{c|}{\textbf{$G_c = G$}} & \multicolumn{2}{c|}{\textbf{reduced $G_c = \lfloor\frac{1}{4} G\rfloor$}} &         \\
     \midrule
$N$ & $G_c$ & fps & $G_c$ & fps & $S$\\
\midrule
128 & 384 & 18.1 & 96 & 25.8 & 1.43 \\
144 & 432 & 10.1 & 108 & 13.0 & 1.29 \\
160 & 486 & 11.9 & 121 & 17.9 & 1.50 \\
170 & 512 & 13.4 & 128 & 20.1 & 1.50 \\
256 & 784 & 6.1 & 196 & 9.7 & 1.59 \\
\bottomrule
\end{tabularx}
\label{tbl:tiny}
\end{table}

\subsection{High-level Parallelization with Autotuning}

In parallel MRI, the reconstruction problem can be partitioned across the domain of multiple receive channels used for signal reception \cite{schaetz2012}. The measured data $y_j$,  corresponding coil sensitivity maps $c_j$ and associated intermediate variables can be assigned to different accelerators $A$. The image $\rho$ has to be duplicated on all accelerators. The sum $\sum_j^J c_j^{\star} t_j$ is partitioned across accelerators according to 
\begin{equation}
\sum_j^J c^{\star}_j t_j = \sum_{a=1}^A \sum_{j \in J_a} c^{\star}_j t_j
\end{equation}
where $J_a$ is the subset of channels assigned to accelerator $a$. This summation amounts to an all-reduce operation, since all accelerators require the computed updates to $\rho$. An alternative decomposition would require communication between GPUs within the FFT operation which is not feasible for the typical vector dimensions in this problem. The resulting speed-up from this parallel decomposition originates from the fact that the FFT of all channels can be computed in parallel. With only one accelerator all $J$ Fourier transforms have to be computed by a single accelerator with a batch-size equal to $J$. A batched FFT computes multiple FFTs of the same size and type on multiple blocks of memory either in parallel or in a sequential manner. With multiple accelerators $J$ is divided by the number of accelerators $A$. This is illustrated in Table \ref{tbl:channel-decomp-speedup-cause} depicting the computation time of Fourier transforms of $10$ 2D vectors of various side lengths $N$ for different numbers of GPUs.

The parallel efficiency of distributing a batched FFT across devices decreases from two to three and four accelerators, because an accelerator can compute multiple 2D FFTs at the same time. In addition, the NLINV implementation uses 10 compressed channels which does not divide without remainder by 3 and 4. 

\begin{table}
\caption{Computation time $t$ in $\mu s$ and  parallel efficiency $E$ of the Fourier transform of $10$ 2D vectors of side length $N$ calculated using the cuFFT batched mode.}
\begin{tabularx}{\columnwidth}{Y|c|cc|cc|cc}
\toprule
      & 1 GPU & \multicolumn{2}{c|}{2 GPUs} & \multicolumn{2}{c|}{3 GPUs} & \multicolumn{2}{c}{4 GPUs}     \\
\midrule
 $N$  & $t$ & $t$ & $E$ & $t$ & $E$ & $t$ & $E$ \\
\midrule
384 & 429   & 230    & 0.93   & 189    & 0.76   & 152    & 0.71   \\
432 & 532   & 280    & 0.95   & 228    & 0.78   & 178    & 0.75   \\
486 & 730   & 387    & 0.94   & 320    & 0.76   & 251    & 0.73   \\
512 & 555   & 288    & 0.96   & 233    & 0.79   & 179    & 0.78   \\
784 & 1699  & 865    & 0.98   & 697    & 0.81   & 529    & 0.80  \\
\bottomrule
\end{tabularx}
\label{tbl:channel-decomp-speedup-cause}
\end{table}

Furthermore, the speed-up is reduced by the communication overhead of calculating $\sum^{A}_{a=1}$
which increases with the number of accelerators. Table \ref{tbl:channel-decomp} shows the reconstruction speed and speed-up for differently sized datasets and varying number of accelerators. To reduce the communication overhead, a peer-to-peer communication technique is employed that allows accelerators to directly access memory of neighbouring accelerators. This is only possible, if accelerators share a PCIe domain. The Supermicro system has 2 PCIe domains that each connect 4 accelerators. The maximum effective number of accelerators that different channels are assigned to is therefore $4$.

\begin{table}
\caption{Image reconstruction speed (fps),  relative speed-up $S_{rel}$ and  overall speed-up $S_{ov}$
for different numbers of GPUs and differently sized datasets.}
\begin{tabularx}{\columnwidth}{c|Y|rr|rr|rr|r}
\toprule
& 1 & \multicolumn{2}{c|}{2 GPUs} & \multicolumn{2}{c|}{3 GPUs} & \multicolumn{2}{c|}{4 GPUs} &     \\
\midrule
  $N$  & fps & fps & $S_{rel}$ & fps & $S_{rel}$ & fps & $S_{rel}$ & $S_{ov}$ \\
    \midrule
128 & 8.3 & 13.1 & 1.6 & 13.5 & 1.0 & 13.8 & 1.0 & 1.7 \\
144 & 2.5 & 3.9  & 1.6 & 4.1  & 1.0 & 4.2  & 1.0 & 1.7 \\
160 & 5.0 & 8.3  & 1.6 & 8.7  & 1.1 & 9.1  & 1.0 & 1.8 \\
170 & 6.0 & 9.6  & 1.6 & 10.1 & 1.0 & 10.1 & 1.0 & 1.7 \\
256 & 2.4 & 3.8  & 1.6 & 4.5  & 1.2 & 4.7  & 1.1 & 2.0 \\
\bottomrule
\end{tabularx}
\label{tbl:channel-decomp}
\end{table}

\label{subsec:temp_decomp}

Because the parallel efficiency of the channel decomposition is limited, investigations focused on decomposing the problem along the temporal domain which results in reconstructions of multiple frames at the same time. The standard formulation of the NLINV algorithm prohibits a problem decomposition along the temporal domain. Frame $n$ must strictly follow frame $n-1$ as $x_{n-1}$ serves as starting and iterative regularization value for $x_{n}$. While maintaining the necessary temporal order, a slight relaxation of the temporal regularization constraint allows for the reconstruction of multiple frames at the same time. The following scheme ensures that the difference in the results of in-order and out-of order image reconstruction remains minimal. The first frame $n=0$ is defined with $\rho$ set to unity and $c_j$ set to zero. The function $h$ maps frame $n$ for each $n>0$ and Newton step $m$ to initialization an regularizations values $x_{h(n,m)}$:

\begin{equation}
h(n, m) = \left\{
  \begin{array}{ll}
    n-1&: n \leq l \vee m = M-1\\
    \lbrack n-o, n-1 \rbrack&: n > l \wedge m < M-1 \\
  \end{array}
\right.
\end{equation}

\begin{figure}
\centering
	\includegraphics[width=80mm]{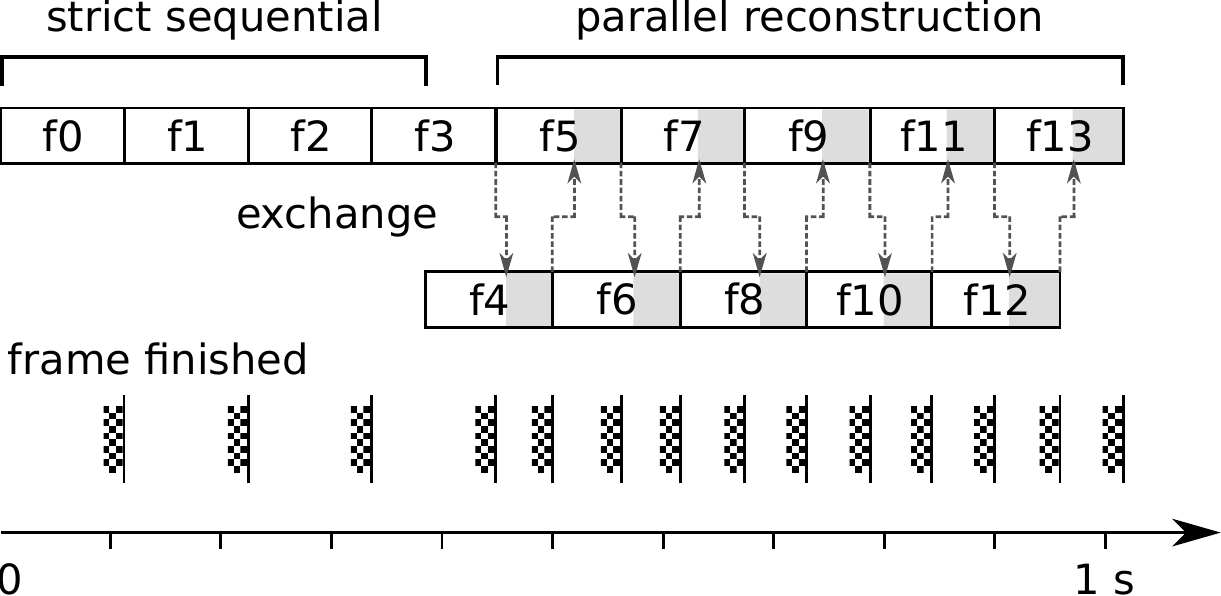} 
	\caption{Temporal decomposition when reconstructing 14 frames. The first 4 frames are reconstructed in strict sequential order, while following images are reconstructed by two threads in parallel. Data is exchanged between threads for the last iteration (i.e., Newton step, grey segment) of each frame. This last iterative calculation takes longest as the CG algorithm requires more time to solve the system of equations. The frame rate is almost doubled by parallel reconstruction in this example.}
	\label{fig:temporal_decomp}
\end{figure}

Due to the complementary data acquisition pattern for sequential frames, the initial frames are of poor quality. The first $l$ images of a series are thus reconstructed in a strict in-order sequence. This allows the algorithm to reach the best image quality in the shortest amount of time. From frame $l+1$ on, frames may be reconstructed in parallel. Initialization and regularization values are chosen to be the most recent available frame within the range $[n-o, n-1]$. The only exception are regularization values in the last Newton step $m = M-1$ where $x_{n-1}$ must be used for regularization. The algorithm thus waits for frame $n-1$ to be computed before proceeding to the last Newton step (compare Figure \ref{fig:temporal_decomp}). Experimental validation shows that the best match of in-order and out-of-order processing while maintaining a speed-up can be achieved by setting $l$ to the number of turns and $o$ to roughly half the number of turns in the interleaved sampling scheme.

%\subsubsection{Autotuning}

Channel decomposition and temporal decomposition can be applied for different imaging situations in different ways. Depending on the acquisition and reconstruction parameters the number of reconstruction threads $T$ (temporal decomposition) and the number of accelerators per reconstruction thread $A$ (channel decomposition) can be adjusted. A prerequisite to this method is that all $(T,A)$ settings yield the same image quality. The parameters $P_{acqu}$ and $P_{reco}$ that have the most impact on image reconstruction speed include
\begin{itemize}
\item the imaging mode: single-slice anatomy, multi-slice anatomy, phase-contrast flow,
\item the data size which depends on the field-of-view and the chosen resolution,
\item the number of frames acquired and
\item the number of virtual channels used.
\end{itemize}
The number of parameters and the ever evolving measurement protocols in a research setting make it difficult to come up with a model that maps the set of acquisition and reconstruction parameters to the optimal set of parallelization parameters $(P_{acqu}, P_{reco}) \rightarrow (T, A)$. Therefore, an autotuning mechanism is employed: The algorithm measures its own run time $R$ and stores its performance along with all acquisition, reconstruction and parallelization parameters in a database $(P_{acqu}, P_{reco}) \rightarrow (T, A) \rightarrow R$. For a given set of $(P_{acqu}$ and $P_{reco})$, the autotuning mechanism can select all recorded run times $R$, sort them and select the set of parallelization parameters $(T,A)$ that yields the best performance. 

The autotuning has an optional learning mode to populate the database with varying parallelization parameters $(T, A)$. If the learning mode is active, the algorithm searches the database for matching performance data and chooses parallelization arguments that do not yet exist in the database. The search space for $(T,A)$ is limited - there are only 16 sets of arguments for the 8-fold GPU reconstruction system used here. The reason for this is the restriction of the channel decomposition stage to the size of the PCIe domain due to peer-to-peer memory access. If $A=1$ then $T=[1,2,3,4,5,6,7,8]$, if $A = 2$, then $T=[1,2,3,4]$ and if $A=[3,4]$ then $T=[1,2]$. 
Table \ref{tbl:autocalibration} shows an excerpt from the autotuning database.

In a clinical setting, all relevant protocols may undergo a learning phase populating the database and covering the entire search space. This can be done in a setup phase  to ensure optimal run times for all clinical scans. The algorithm is also capable of sorting acquisition and reconstruction parameters. This allows the autotuning mechanism to find good $(T,A)$ parameters for new measurement protocols it has never seen before.

\begin{table}
\caption{Autotuning for reconstructing single-slice, dual-slice and phase-contrast flow MRI acquisitions. $N$ = $160$ for all datasets results in an grid size of $486^2$.}
\begin{tabularx}{\columnwidth}{Y|ccr|ccr}
\toprule
\multicolumn{7}{c}{\textbf{single-slice anatomic MRI}}                        \\ 
\midrule
      & \multicolumn{3}{c|}{worst configuration}        & \multicolumn{3}{c}{best configuration}      \\ 
\midrule
frames     & threads      & GPUs/thread   & fps    & threads      & GPUs/thread & fps  \\
5    & 2      & 4  & 1.9   & 1      & 2   & 3.7    \\
10    & 2      & 4  & 3.0  & 1      & 2   & 5.0    \\
25    & 1      & 1  & 4.7   & 2      & 2   & 7.7   \\
50    & 1      & 1  & 4.9   & 3      & 2   & 11.0  \\
200   & 1      & 1  & 4.9  & 3      & 2   & 18.1   \\ 
\midrule
\multicolumn{7}{c}{\textbf{dual-slice anatomic MRI}} \\ 
\midrule
      & \multicolumn{3}{c|}{worst configuration}        & \multicolumn{3}{c}{best configuration}      \\ 
\midrule

frames     & threads      & GPUs/thread   & fps    & threads      & GPUs/thread  & fps  \\
5     & 2      & 4  & 3.2   & 2      & 1   & 5.9    \\
10    & 1      & 1  & 4.6   & 2      & 2   & 8.0    \\
25     & 1      & 1  & 5.0  & 4      & 2   & 12.3  \\
50     & 1      & 1  & 5.1   & 4      & 2   & 18.4 \\
200   & 1      & 1  & 5.1   & 4      & 2   & 28.1  \\ 
\midrule
\multicolumn{7}{c}{\textbf{phase-contrast flow MRI}}                 \\ 
\midrule
      & \multicolumn{3}{c|}{worst configuration}        & \multicolumn{3}{c}{best configuration}      \\ 
\midrule

frames     & threads      & GPUs/thread   & fps    & threads      & GPUs/thread  & fps  \\
5     & 2      & 4  & 1.6  & 2      & 1   & 2.6    \\
10    & 1      & 1  & 1.8  & 2      & 2   & 3.6    \\
25      & 1      & 1  & 1.9  & 4      & 2   & 5.8  \\
50     & 1      & 1  & 1.9  & 4      & 2   & 7.5   \\
200    & 1      & 1  & 1.9 & 4      & 2   & 10.7  \\
\bottomrule
\end{tabularx}
\label{tbl:autocalibration}
\end{table}

\section{Discussion}

This work describes the successful development of a highly parallelized implementation of the NLINV algorithm for real-time MRI where dynamic image series are reconstructed, displayed and stored with little or almost no delay. The computationally demanding iterative algorithm for dynamic MRI was decomposed and mapped to massive parallel hardware. This implementation is deployed to several research groups and represents a cornerstone for the evaluation of the clinical relevance of a variety of real-time MRI applications, mainly in the field of cardiac MRI \cite{zhang2010cmri,zhang2014real}, quantitative phase-contrast flow MRI \cite{joseph2012flow,untenberger2015advances}, and novel fields such as oropharyngeal functions during swallowing \cite{zhang2015diagnosis}, speaking \cite{niebergall2013real}, and brass playing \cite{iltis2015real}.

As examples, supplementary videos 1 and 2 show T1-weighted radial FLASH MRI acquisitions at 33.32 ms acquisition time (TR = 1.96 ms, 17 radial spokes covering k-space) of a human heart in a short-axis view and midsagittal tongue movements of an elite horn player, respectively. The cardiac example employed a {256$\times$256} mm$^3$ field-of-view, 1.6 mm in-plane resolution (6 mm slice thickness), and $N=160$ data samples per spoke (i.e., grid size) which resulted in a slightly delayed reconstruction speed of about 22 fps vs 30 fps acquisition speed. In contrast, the horn study yielded real-time reconstruction speed of 30 fps because of a slightly smaller {192$\times$192} mm$^3$ field-of-view at 1.4 mm in-plane resolution (8 mm slice thickness), and only $N=136$ samples per spoke.

Some of the optimization methods discussed here are specific to real-time MRI algorithms, but the majority of techniques translate well to other MRI applications or even to other medical imaging modalities. Functional decomposition can be employed in any signal or image processing pipeline that is built on multiple processing units like multi-core CPUs, heterogeneous systems or distributed systems comprised of FPGAs, digital signal processors, microprocessors, ASICs, etc. Autotuning is another generic technique that can be useful in many other applications: if an algorithm implementation exposes parallelization parameters (number of threads, number of GPUs, distribution ratios etc.), it can be tuned for each imaging scenario and each platform individually. The gridsize optimization technique can also be universally employed: if an algorithm allows choosing the data size (within certain boundaries) due to re-gridding or interpolation, the grid size should be chosen such that following processing steps exhibit optimal performance. This is valuable, if following processing steps do not exhibit linear performance such as for example specific FFT implementations. 

\section{Conclusion}

Novel parallel decomposition methods are presented which are applicable to many
iterative algorithms for dynamic MRI. Using these methods to parallelize the NLINV
algorithm on multiple GPUs it is possible
to achieve online image reconstruction with high frame rates. For
suitable parameters choices, real-time reconstruction can be achieved.

\section*{Conflict of Interest} 

Martin Uecker and Jens Frahm hold a patent about the real-time MRI acquisition and reconstruction technique discussed here.
All other authors declare that they have no competing interest.

\section*{Acknowledgement}
{We acknowledge support by the German Research Foundation and the Open Access Publication Funds of the G\"ottingen University.}

\bibliographystyle{ieeetr} 
\bibliography{compmri}

\end{document}